\documentclass[a4paper,11pt]{article}
\usepackage{rsfs}
\usepackage{latexsym}
\usepackage{amssymb}
\usepackage{graphicx}
\begin{document}

\title{{\Large \bf  Electromagnetic Form Factors of Nucleons with QCD 
Constraints}\\
 {\bf{\large Sytematic Study of the Space and Time-like Regions}}\\ 
 \vspace{24pt}}
\author{
 Susumu FURUICHI\footnote{Present address: Sengencho 3-2-6, Higashikurume Tokyo 203-0012 }\\
 Department of Physics, Rikkyo University, Toshima Tokyo 171-8501, Japan 
 {}\\
 {}\\
 Hirohisa ISHIKAWA\footnote{e-mail address: ishikawa@meikai.ac.jp}\\
 Department of EconomyCMeikai University, Urayasu Chiba, 279-8550, Japan
{}\\
{}\\
 Keiji WATANABE\footnote{ e-mail address keijiwatanabe888@yahoo.co.jp; 
 Present address: Akazutumi, 5-36-2, 
 Setagaya Tokyo 156-0044}\\
 Department of Physics, Meisei University, Hino Tokyo 191-8506, Japan}
\date{}
\maketitle
\begin{abstract}
  Elastic electromagnetic form factors of nucleons are investigated
 both for
   the time-like and the space-like momentums under the condition that the QCD 
   constraints are satisfied asymptotically. The unsubtracted dispersion 
  relation with the superconvergence conditions are used as a realization of 
  the QCD conditions. The experimental data are analyzed by using the 
  dispersion formula and it is shown that the calculated form factors
   reproduce the experimental data reasonably well.
\end{abstract}
\renewcommand{\baselinestretch}{1.5}

\section{Introduction}
We have investigated the nucleon electromagnetic form factors by 
using the dispersion relation, which 
 worked in understanding the low energy data very well. We were able to
  realize the  
low energy experimental data for the space-like momentum transfer 
\cite{fkw}-\cite{fw2}.
For the low momentum, the vector dominance model is qualitatively 
 valid in explaining the electromagnetic form factors of nucleons  
except for the $\rho$ meson mass, which should be taken much smaller than 
the experimental value. The problem of the $\rho$ meson mass was solved by
  taking into account the uncorrelated two pion contribution. The dispersion 
 relation turned out to be very effective for this purpose. 
 
 Phenomenologicaly, the magnetic and electric form factors of nucleons, 
 $G_M^N(t)$ and $G_E^N(t)$ respectively, 
  are proportional to the dipole formula 
  $G_D(t)=1/(1+|t|/0.71)^2$, where $t$ is the squared space-like momentum 
  transfer expressed in the unit of $({\rm GeV}/c)^2$. \footnote{We adopt the convention $t<0$ for the space-like and 
  $t>0$ for the time-like momentum. We write $t=-Q^2$ for the 
  space-like momentum and $t=s>0$ for the time-like momentum.} The dipole 
  formula represents the experimental data of form factors fairly well 
for large range of momentum. 

To be precise, the experimental data, however, decrease more rapidly than the
  dipole formula. This is compatible with the prediction of 
 perturbative QCD
  (PQCD), where the elastic form factors of
  hardons decrease asymptotically for large squared momentum as compared with
   the  dipole formula \cite{brodsky}: For the boson form
 factors, $F(t)\to {\rm const}/\ln |t|$  for $|t|\to\infty$ and for the nucleon
 form factors, $F_1^N$ and $F_2^N$, the charge and magnetic moment form 
 factors, respectively, decrease as $F_1^N(t)\to 
 {\rm const}\,t^{-2}(\ln|t|)^{-\gamma}$
  and $F_2^N(t) \to {\rm const}\,t^{-3}(\ln|t|)^{-\gamma}$ with $\gamma$ 
  ($\ge2$) being a constant. Consequently, $G_M^N$ and $G_E^N$ decrease as 
  $t^{-2}(\ln |t|)^{-\gamma}$.
 
  We showed that the QCD conditions are incorporated by assuming 
  superconvergent dispersion relation (SCDR) for the form factors and
  investigated the pion and kaon electromagnetic form factors. It was shown
   that the experimental data are
  reproduced both for the space-like and the time-like momentum
   \cite{nw1,nw2,inw}. 
 
 Different from the boson form factors, for the nucleons the unphysical regions  of absorptive 
 parts are not observed for $s<4m^2$, with $m$ being the nucleon mass. 
 This makes the problem of the nucleon 
 form factors difficult, as was observed by R. Wilson in his 
 review article \cite{wilson}, in which he emphasized the importance of
  systematic study of
  the time-like and space-like regions for the nucleon form factors. 
 
 It is the purpose of this paper to examine the nucleon electromagnetic 
 form factors to see if it is possible to realize the experimental data 
 with the QCD constraints satisfied, where the space-like and time-like 
 regions are treated on equal footing in the chi square analysis. 

 Organization of this paper is given as follows: In Sec.2 we summarize on 
 the dispersion relation for the nucleon electromagnetic form factors 
  with the QCD condition imposed. 
  In Sec.3, imaginary parts of 
the form factors are given for the low, intermediate and asymptotic regions.
 In Sec.4 remarks conserning numerical analysis are given and the numerical results are summarized in Sec.5. The final section is devoted to general
  discussions.
 
 \section{Dispersion relation for the nucleon form factors}

As is mentioned in Sec.1, the electromagnetic form factors approach zero 
asymptotically for $t \to \infty$. Therefore, we may assume 
the unsubtracted dispersion relations for the charge and magnetic moment form 
factors $F_1^{I}$ and $F_2^I$, respectively, with $I$ 
denoting the isospin state $I=0,1$. That is,
\begin{equation}
 F_i^{I}=\frac{1}{\pi}\int_{t_0}^{\infty}dt'
  \frac{{\rm{Im}}\, F_i^{I}(t')}{t'-t}, \label{unsubtracted}
\end{equation}
where the threshold is $t_0=4\mu^2$. Here $\mu$ is the pion mass being  
taken as the average of the neutral and charged pions.

We briefly summarize the asymptotic
 theorems which are used to incorporate the 
 constraints of PQCD \cite{brodsky}, where the proof is given in 
 Ref.\cite{nw1}. 
Let $F(t)$  satisfy the dispersion relation (\ref{unsubtracted}), 
and ${\rm Im}F$ be given as
\begin{equation}
 {\rm Im}F(t') = \frac{c}{[\ln (t'/Q_0^2)]^{\gamma}}
   +O\left(\frac{1}{[\ln(t'/Q_0^2)]^{\gamma+1}}\right)
\end{equation}
for $t\to \infty$ with $\gamma>1$. Then $F(t)$  approaches
\begin{equation}
 F(t)=\frac{1}{\pi}\int_{t_0}^{\infty}dt'
 \frac{c}{(t'-t)[\ln (t'/Q_0^2)]^{\gamma}}\to 
 \frac{c}{\pi(\gamma-1)\ln(|t|/Q_0^2)]^{\gamma-1}}  \label{disp1}
\end{equation}
for $t\to \pm \infty$. 

Generally, if ${\rm Im}F(t)$ tends to
\begin{equation}
t^{\prime\,n+1}{\rm Im}F(t')\,\,\to \frac{c}{[\ln(t'/Q_0^2)]^{\gamma}}
  +O\left(\frac{1}{[\ln(t'/Q_0^2)]^{\gamma+1}}\right) \label{asymp2}
\end{equation}
for $t \to \pm \infty$ and the superconvergence conditions
\begin{equation}
 \int_{t_0}^{\infty}dt't^{\prime k}{\rm Im}F(t')=0,\quad k=0,1,\cdots,n,
  \label{asym4}
\end{equation}
are satisfied, $F(t)$ given by (\ref{unsubtracted}) approaches for $t \to \pm\infty$ 
\begin{equation}
 F(t)= \frac{1}{\pi}\int_{t_0}^{\infty}dt'\frac{{\rm Im}F(t')}{t'-t}
     \to \frac{1}{t^{n+1}}\frac{c}{\pi(\gamma-1)[\ln(|t|/Q_0^2)},
\end{equation}
which can be proved by using (\ref{disp1}), (\ref{asymp2}) and (\ref{asym4}) 
together with the identity
$$
 \frac{1}{t'-t}=-\frac{1}{t}\Big\{1+\frac{t'}{t}+\cdots
   +\Big(\frac{t'}{t}\Big)^n\Big\}+\frac{1}{t^{n+1}}
   \frac{t^{\prime n+1}}{t'-t}.
$$

 The QCD constraints for the nucleon 
form factors are, therefore, attained by assuming the unsubtracted dispersion 
relation and 
 the superconvergence conditions for ${\rm{Im}}\,F_i^I$
\begin{eqnarray}
 && \frac{1}{\pi}\int_{t_0}^{\infty}dt'\,{\rm Im}F_1^I(t')
     =\frac{1}{\pi}\int_{t_0}^{\infty}dt't'\,{\rm Im}F_1^I(t')=0, 
                                                       \label{sup1}\\ 
 && \frac{1}{\pi}\int_{t_0}^{\infty}dt'\,{\rm Im}F_2^I(t')
     =\frac{1}{\pi}\int_{t_0}^{\infty}dt't'\,{\rm Im}F_2^I(t') \nonumber \\
 &&\qquad\qquad\qquad\qquad\,\,\,
     =\frac{1}{\pi}\int_{t_0}^{\infty}dt't^{\prime\,2}\,{\rm Im}F_2^I(t')=0,
                                                       \label{sup2}
\end{eqnarray}
where ${\rm Im}F_i^{I}(t')$ satisfies the asymptotic conditions for $t' \to
 \infty$
\begin{equation}
 t^{\prime\, i}{\rm Im}F_i^{I}(t') \to {\rm const}/[\ln (t'/Q_0^2)]^{\gamma}
 \quad (i=1,2).   \label{asymptotic}
\end{equation}
$Q_0$ and $\gamma(\ge2)$ are constants, the latter of which is written in terms
of the anomalous dimension of the renormalization group in QCD.

In addition to the conditions (\ref{sup1}) and (\ref{sup2}) we impose the 
normalization conditions at $t=0$:
\begin{eqnarray}
 \frac{1}{2}&=&\frac{1}{\pi}\int_{t_0}^{\infty}dt'\,{\rm Im}F_1^{I}(t')/t', 
                                    \label{norm1} \\
 g^I &=& \frac{1}{\pi}\int_{t_0}^{\infty}dt'\,{\rm Im}F_2^{I}(t')/t',
                                    \label{norm2}
\end{eqnarray}
where $g^I$ are the anomalous moments of nucleons with the isospin $I$.
 
 \section{Imaginary part of the form factors}

Let us discuss the imaginary parts of nucleon form factors, which are broken 
up into three parts:
 The low momentum, the intermediate, and the asymptotic regions.

 \subsection{Low momentum region}
 
The imaginary parts of the charge and magnetic moment form factors, 
${\rm Im}F_i^{V}$, are given in terms two pion contributions
\begin{eqnarray}
 {\rm Im}[F_1^{V}(t)/e] &=& \frac{m}{2}\frac{(t-4\mu^2)}{4m^2-t}
  \left(\frac{t-4\mu^2}{t}\right)^{1/2}  \nonumber \\
  &&\times {\rm Re}\Big[M^{*}(t)
  \Big\{f_{+}^{(-)1}(t)
  -\frac{t}{4m^2}\frac{m}{\sqrt 2}f_{-}^{(-)1}(t)\Big\}\Big], \\
 {\rm Im}[2mF_2^{V}(t)/e] &=& \frac{m}{2}\frac{(t-4\mu^2)}{(4m^2-t)}
  \left(\frac{t-4\mu^2}{t}\right)^{1/2}                  \nonumber\\
   &&\times {\rm Re}\Big[M^{*}\Big\{\frac{m}{\sqrt 2}f_{-}^{(-)1}(t)
      -f_{+}^{(-)1}(t)\Big\}\Big], \label{ImH}
\end{eqnarray}
where $f_{\pm}^{(-)1}(t)$ are helicty
 amplitudes for $ \pi\pi\leftrightarrow N\bar{N}$ , $M(t)$ is the pion 
 form factor and $\mu$ is the pion mass.  
For the helicity amplitudes we use the numerical values given by H\"ohler and 
Schopper \cite{hoe} and parameterize $M(t)$ according to them.
\begin{equation}
 M(t) = 
  t_{\rho}\{1+(\Gamma_{\rho}/m_{\rho}d)\}
  [t_{\rho}-t-im_{\rho}^2\Gamma_{\rho}(q_t/q_{\rho})^3\sqrt{t}]^{-1},
\end{equation}
where $m_{\rho}$ and $\Gamma_{\rho}$ are the $\rho$ meson mass and width 
respectively and
\begin{equation}
 t_{\rho}=m_{\rho}^2,\quad q_{\rho}=\sqrt{t_{\rho}-\mu^2},\quad
  d=\frac{3\mu^2}{\pi t_{\rho}}\ln\frac{m_{\rho}+2q_{\rho}}{2\mu}
   +\frac{m_{\rho}}{2\pi q_{\rho}}\Big(1-\frac{2\mu^2}{t_{\rho}}\Big).
\end{equation}
The imaginary parts thus obtained are denoted as ${\rm Im}F_i^{H}\,\,(i=1,2)$ 
hereafter. It must be remarked that the $\rho$ meson contribution is included   in the helicity amplitudes of Ref.\cite{hoe}.

\subsection{Intermediate region}

The intermediate states $4\mu^2\le t \le \Lambda_1^2$ are approximated 
by the addition of the Breit-Wigner terms. with the imaginary part 
  parameterized as follow:
\begin{equation}
 {\rm Im}f_R^{BW}(t) = \frac{g}{(t-M_R^2)^2+g^2},
\end{equation}
 where
\begin{equation}
g=\frac{\Gamma M_R^2(M_R^2+t_{res})^3}{t_{res}^2(M_R^2-t_0)^{3/2}}
 \sqrt{\frac{(t-t_0)^3}{t}}\frac{t^2}{(t+t_{res})^3},
\end{equation}
where $M_R$ and  $\Gamma$ are the mass and width of resonance, respectively. 
$t_0$ is the threshold $t_0=4\mu^2$ and $t_{res}$, being treated as an 
adjustable parameter, is introduced to cut-off the Breit-Wigner
 formula.

We write the intermediate part as the summation of resonances
\begin{equation}
 {\rm Im} F_i^{BW,I}=\sum_n a_n^{I,i}f_{nR}^{I}, \label{ImBW}
\end{equation}
where $I$ is the isospin and $n$ is the labeling of resonances (see Table I).
Here the suffix $i$ denotes $i=1,2$, corresponding to the charge and 
magnetic moment form factors $F_1^N$ and $F_2^N$. The same formulas for
 $f_{nR}^{I}$ are used for $i=1$ and $i=2$.

 \subsection{Asymptotic region}
 
 To calculate the absorptive part of form factors for the asymptotic region,
  we need the running coupling constant $\alpha$ for the time-like momentum. 
  We perform the analytic continuation to the time-like momentum by assuming 
 the dispersion relation 
 for $\alpha$; application of the so called analytic regularization 
 \cite{jones}-\cite{dok2}.  

 Let $\alpha_S(Q^2)$ be the running coupling constant calculated by the 
perturbative QCD as a function of the squared momentum $Q^2$ for the
space-like momentum expressed in the Pade form. 
\begin{eqnarray}
 \alpha_S(Q^2) &=& \frac{4\pi}{\beta_0}\Big[\ln(Q^2/\Lambda^2)
  +a_1\ln\{\ln(Q^2/\Lambda^2)\} \nonumber \\
  &&+a_2\frac{\ln\{\ln(Q^2/\Lambda^2)\}}{\ln(Q^2/\Lambda^2)}
    +\frac{a_3}{\ln(Q^2/\Lambda^2)}+\cdots\Big]^{-1}.  \label{PQCD}
\end{eqnarray}
Here $\Lambda$ is the QCD scale parameter, and $a_i$ are given in terms of
 the $\beta$ function of QCD,
\begin{equation}
 a_1=2\beta_1/\beta_0^2, \quad a_2=4\frac{\beta_1^2}{\beta_0^4}, \quad 
 a_3=
  \frac{4\beta_1^2}{\beta_0^4}\left(1-\frac{\beta_0\beta_2}{8\beta_1^2}\right),
\end{equation}
where
\begin{equation}
 \beta_0=11-\frac{2n_f}{3},\quad \beta_1=51-\frac{19n_f}{3},\quad 
 \beta_2=2357-\frac{5033}{9}n_f+\frac{325}{27}n_f^2
\end{equation}
with $n_f$ being the number of flavor.
We perform the analytic continuation of the squared momentum $Q^2$ to the 
time-like region, $s$, by the replacement in (\ref{PQCD})
\begin{equation}
 Q^2\to e^{-i\pi}s.
\end{equation}
Then the effective coupling constant becomes complex; 
 $\alpha_S(q)={\rm Re}[\,\alpha_S(s)]+i{\rm Im}[\,\alpha_S(s)]$
with
\begin{eqnarray}
 {\rm Re}[\,\alpha_S(s)]&=&\frac{4\pi u}{\beta_0 D(s)}, \\
 {\rm Im}[\,\alpha_S(s)]&=&\frac{4\pi v}{\beta_0 D(s)}.
\end{eqnarray}
We write
\begin{eqnarray*}
 \alpha_S(s) &=& 1/(u-iv)=\frac{u+iv}{D}, \\
          D  &=& u^2+v^2,
\end{eqnarray*}
where
\begin{eqnarray}
 u &=& \ln(s/\Lambda^2)+\frac{a_1}{2}\ln\{\ln^2(s/\Lambda^2)+\pi^2\} 
                                               \nonumber\\
     && +\frac{a_2}{\ln^2(s/\Lambda^2)+\pi^2}
         \Big[\frac{1}{2}\ln(s/\Lambda^2)\ln\{\ln^2(s/\Lambda^2)
       +\pi\theta\}\Big]                   \nonumber \\
     &&+\frac{a_3\ln(s/\Lambda^2)}{\ln^2(s/\Lambda^2)+\pi^2},  \\
 v &=& \pi+a_1\theta                       \nonumber \\
     &&-\frac{a_2}{\ln^2(s/\Lambda^2)+\pi^2}
        \Big[\frac{\pi}{2}\ln\{\ln^2(s/\Lambda^2)+\pi^2\}
        -\theta\ln(s/\Lambda^2)\Big]       \nonumber \\
     &&-\frac{\pi a_3}{\ln^2(s/\Lambda^2)+\pi^2},
\end{eqnarray}
and
\begin{equation}
 \theta=\tan^{-1}\{\pi/\ln(s/\Lambda^2)\}.
\end{equation}
The running coupling constant is given by the dispersion integral 
 both for the space-like and the time-like momentum
\begin{equation}
 \alpha_R(t)=\int_{Q_0^2}^{\infty}dt'\frac{\sigma(t')}{t'-t} \label{regular}
\end{equation}
with
\begin{equation}
 \sigma(t')=4\pi v/\beta_0D.
\end{equation}
$\alpha_R(t)$ represented by (\ref{regular}) is called analytically 
regularized running coupling constant as it has no singular point for $t<0$. 
The regularization eliminates the ghost 
pole of $\alpha_S(Q^2)$ appearing  at the point
\begin{equation}
 Q^2=Q^{*2}=\Lambda e^{u^{*}},
\end{equation}
where $u^{*}=0.7659596\cdots$ for the number of flavor $n_f=3$.
Calculating (\ref{regular}), we find that $\alpha_R(t)$ is approximately given by the simple formula with the ghost pole subtracted
\begin{equation}
 \alpha_R(t)\approx \alpha_S(Q^2)-A^{*}/(Q^2-Q^{*2})   \label{regular-1},
\end{equation}
where the residue $A^{*}$ is
\begin{equation}
 A^{*}=4\pi\Lambda^2e^{u^{*}}/\Big\{\beta_0
  \Big(1+\frac{a_1}{u^{*}}
  -a_2\frac{\ln u^{*}}{u^{*2}}+\frac{a_2-a_1}{u^{*2}}\Big)\Big\}.
\end{equation}
We use (\ref{regular-1}) as the regularized coupling constant; for the 
time-like momentum we replace $Q^2=e^{-i\pi}s$ as was mentioned before.

The QCD parts, $F_i^{QCD,\,I}$ ($i=1,2$; $I$ = 0,1) are written as follows:
\begin{equation}
 F_i^{QCD,\,I}(t)= \hat{F}_i^{QCD,\,I}(t)h_i(t),  \label{qcd1}
\end{equation}
where $\hat{F}_i^{QCD,\,I}$'s are given as expansion in terms of the 
running coupling constant 
\begin{equation}
 \hat{F}_i^{QCD,\,I}(t)=\sum_{j\ge 2}c_j^{QCD,\,I}
  \{\alpha_R(t)\}^j
\end{equation}
for the space-like momentum $(t<0)$. We multiply the function $h_i(t)$ in 
(\ref{qcd1}) to assure the convergence of the superconvergence 
conditions (\ref{sup1}) and (\ref{sup2}). The following formula is assumed for 
$h_i(t)$:
\begin{equation}
 h_i(t)=\left(\frac{t-t_{Q}}{t+t_1}\right)^{3/2}
       \left(\frac{t_2}{{t+t_2}}\right)^{i+1}\quad (i=1,2),
\end{equation}
which may be interpreted as the form factor for $\gamma \to q\bar{q}$ with 
$t_Q$ being the threshold of the quark antiquark pair.
The parameters $t_Q$, $t_1$ and $t_2$ are 
taken as adjustable parameters and will be determined by the analysis of 
experimental data.

For the time-like momentum ($t>0$), we perform the analytic 
continuation of the regularized effective coupling constant $\alpha_R(t)$ 
to $\alpha_R(s)$ through the equation
\begin{equation}
 \alpha_R(s)=\alpha_R(Q^2e^{-i\pi})={\rm Re}[\alpha_R(s)]
    +i{\rm Im}[\alpha(s)].
\end{equation}
We take three loop approximation for the effective coupling constant and 
express the QCD part as follows:
\begin{equation}
 \hat{F}_i^{QCD,I}(s)=\sum_{2\le j \le 4}c_{i,j}^{QCD,\,{\rm I}}
  \{\alpha_R(t)\}^j.  \label{cQCD}
\end{equation}
The summation in (\ref{cQCD}) begins in the second order in the effective
coupling constant so as to realize the logarithmic decrease of the nucleon 
form factors.

Imaginary part of (\ref{cQCD}) is obtained to be
\begin{eqnarray}
 {{\rm Im}\hat{F}}_i^{QCD,I} &=& 2c_{i,2}^{QCD,\,{\rm I}}{\rm Re}\,\alpha_R
 {\rm Im}\,\alpha_R+
   c_{i,3}^{QCD,\,{\rm I}}[3({\rm Re}\,\alpha_R)^2{\rm Im}\,\alpha_R
   -({\rm Im}\,\alpha_R)^3] \nonumber \\
   &&+c_{i,4}^{QCD,\,I}[4({\rm Re}\,\alpha_R)^3{\rm Im}\,\alpha_R
     -4{\rm Re}\,\alpha_R({\rm Im}\,\alpha_R)^3]+\cdots, \label{ImQCD}
\end{eqnarray}
and ${\rm Im}F_i^{QCD,\,I}(s)={\rm Im}\hat{F}_i^{QCD\,I}(s)h_i(s)$.

We write the low energy part, intermediate resonance part and asymptotic 
QCD parts of form factors as $F_i^{\rm{ H}}$, $F_i^{BW, I}$ and $F_i^{QCD,\,I}$,  which are 
given by the dispersion integral with the imaginary parts (\ref{ImH}),
 (\ref{ImBW}) and (\ref{ImQCD}), respectively. The form factors $F_i^{I}$ 
 are defined by adding them up. We impose the conditions (\ref{sup1}) and 
 (\ref{sup2}) on ${\rm Im}F_i^{I}$ so that the QCD conditions are satisfied.

 \section{Numerical analysis}

We analyzed the experimental data of nucleon electro-magnetic form factors 
for the space-like momentum $G_M^p/\mu_pG_D$, $G_E^p/G_D$, $G_M^n/\mu_nG_D$, 
$G_E^n$ and the ratio $\mu_p G_E^p/G_M^p$ and for the time-like momentum 
$|G^p|$ and $|G^n|$.  The parameters in the 
form factors are determined so that the calculated results realize the 
experimental data. In addition to the data used in our previous analysis 
\cite{fw2}, \cite{fw3} we used the data in Refs.\cite{bostedGMp}-
\cite{antonelliGntm}. In our analysis we treat both of space-like and 
time-like regions on equal footing in the chi square analysis.\\

\noindent Space-like region:\\

For $G_M^{N}/\mu^{N}$ and $G_E^N$ we used the ratio to the dipole formula $G_D$ as we have done in Refs.\cite{fw1}, \cite{fw2}, \cite{fw3}. 
In addition to them we take into account in the chi square analysis the data of
the ratio $\mu^p G_E^p/G_M^p$ obtained by the polarization experiments. \\

\noindent Time-like region:\\

 Experimentally, the proton and neutron form factors 
$|G^p|$ and $|G^n|$ for the time-like momentum are obtained by using the 
formula for the cross section  $\sigma_0$ for the processes 
$e+\bar{e} \to N+\bar{N}$ or $N+\bar{N} \to e+\bar{e}$, which is given as
\begin{equation}
 \sigma_0=\frac{4\pi\alpha^2\nu}{3s}\left(1+\frac{2m_N^2}{s}\right)
  |G(s)|^2,
  \label{exp}
\end{equation}
where $\alpha$ is the fine structure constant, $m_N$ and $\nu$ are
 the mass and velocity of the nucleon $N$, respectively.  
$|G_M^N|$ are estimated from $|G|$ under the assumption $G_M=G_E$ or $G_E=0$ \cite{ablikimGptm}. $\sigma_0$ is now expressed in terms of 
$G_M^N$ and $G_E^N$
\begin{equation}
 \sigma_0=\frac{4\pi\alpha^2\nu}{3s}
 \left(|G_M^N|^2+\frac{2m_N^2}{s}|G_E^N|^2\right). \label{theory}
\end{equation}
Equating (\ref{exp}) and (\ref{theory}), we have
\begin{equation}
 |G|^2=\frac{|G_M^N|^2+2m_N^2|G_E^N|^2/s}{1+2m_N^2/s}.  \label{timedata}
\end{equation}
Substituting our calculated result of form factors to the right hand side 
of (\ref{timedata}), we 
get the theoretical value for $|G|$, which is compared with the experimental 
data for the magnetic form factor obtained under the assumption $G_M=G_E$.

The parameters appearing in our analysis are the following: \\
Residues at resonances, coefficients appearing in the expansion by the 
QCD effective coupling constants, cut-offs for the low, intermediate and 
asymptotic regions $\Lambda_1,\,\,\Lambda_2$ and $\Lambda_3$, respectively.
In addition to them  we have parameters in the Breit-Wigner formula and 
the convergence factor $h$ of QCD contribution, $t_0$, $t_{res}$, $t_1,\,\,
t_2,\,\, t_3$.

We have taken the masses and the widths of resonances as adjustable parameters.
As the superconvergence constraints impose very stringent conditions on the 
form factors, it was necessary to take the masses and widths as parameters. \\

 \section{Numerical results}

 We give in Table I and II the results for the parameters; 
  in Table I the masses and widths of resonances and residues at 
resonance poles obtained by our analysis and in Table II the coefficients
 $c_{i,j}^{QCD,\,I}$ $(i=1,2; \,\,j=2,3,4;\,\,I=0,1)$ 
 in the expansion in terms of the
 effective coupling constant $\alpha_R$ of QCD defined by (\ref{cQCD}). \\The
  value of 
 $\chi^2$ is obtained to be $\chi^2_{tot}=267.0$, which includes both 
 the data of space-like and time-like regions. The $\chi^2$ value for 
 the space-like data is $\chi^2_{space}=233.3$. 
 The total number of data used
  in the chi square analysis is 209 and the number of parameters is 36. 
  Therefore, $\chi_{tot}^2/Df=1.54$.
 For the time-like momentum, the data of Ablikim et al. \cite{ablikimGptm}
  (2005) is systematically  smaller than that of Antonelli et al.
  \cite{antonelliGntm} (1998).  In the present analysis
  we restricted ourselves to the (2005) data.  We were able to get much 
better result both 
 for the  time-like and for the space-like part than the result obtained by 
 the restriction to (1998) data. 
 
 We summarize the parameters obtained by the chi square analyis in Table I and II. \\

\begin{tabular}{llllll} \hline
isospin $I$&$n$&mass (GeV)&width (GeV)&$\quad a_1^{I,n}$ (GeV$^2$)& 
$\,\,\,\, a_2^{I,n}$ (GeV$^2$)
    \\ \hline 
  {} & 1 & 1.367  & 0.324 & $\,\,\,\,\,$-6.7       &  $\,\,\,\,\,$1.06 \\ 
  {} & 2  & 1.376  & 0.220  & $\,\,\,\,\,$9.562613 & -17.81168 \\ 
  $I=1$& 3  & 1.6096 & 0.26   & $\,\,\,\,$-8.323391  & $\,\,$11.18087  \\ 
  {} & 4  & 1.832  & 0.381   & $\,\,\,\,$ 5.746830      & $\,\,$-5.941032 \\ 
  {} & 5  & 2.320  & 0.430  & $\,\,\,\,$-0.40          & $\,\,$ 0.40   \\  \hline
  {} &1 & 0.78256& 0.844$\times10^{-2}$ & $\,\,\,$0.9123977 
   & $\,\,$ 0.5257854 \\
  {} &2  & 1.01945& 0.426$\times10^{-2}$ & -3.302706 &
   $\,\,\,\,$0.5371316 \\
  $I=0$ &3   & 1.227  & 0.1609 & $\,\,\,$0.6303140$\times10^2$  &
   $\,\,$-1.981666 \\
  {} &4  & 1.472  & 0.2123 & $\,$-1.734902       & $\,\,$-2.589660  \\ 
  {} &5 & 1.530  & 0.1416  & $\,$-2.460598          & $\,\,$ 4.147711  \\ 
  \hline
\end{tabular}\\

  $\quad$ Table I  Parameters obtained by the analysis. Residues at 
  resonances. \\
 {}\\
 
\begin{tabular}{lllll} \hline
isospin $I$& $i$ & $c_{i,2}^{QCD,\,I}$ & $c_{i,3}^{QCD,\,I}$ &
 $c_{i,4}^{QCD,\,I}$
 \\ \hline
 $I=1$ & 1 & -2.403217 & \,\,0.220$\times 10^2$ & \,-4.40  \\
 {}  & 2 & \,\,5.072373 & -0.510$\times 10^2$ & 
 \,\,\,\,0.543$\times 10^2$\\ \hline
 $I=0$ & 1 & \,2.964186  &  -0.195 $\times 10^2$ & \,\,-0.687$\times 10^2$ 
 \\
 {}  & 2 & -6.333821 & \,\,0.770$\times 10^2$ & \,\,-0.1982$\times 10^3$ \\
   \hline
\end{tabular} \\

  Table II Coefficients of expansion in terms of the effective \\
 $ \qquad\qquad\quad$ of coupling constant QCD defined in (\ref{cQCD}).\\

   The parameters $t_1$, $t_2$, $t_{res}$, $t_Q$ and $\Lambda_1$ are determined
    as follows:\\
 $t_1 = 0.2070\times 10^3\,\, ({\rm GeV/c})^2$, 
 $t_2 = 0.2240\times 10^3\,\,({\rm GeV/c})^2$, 
 $t_{res} = 0.2082\times 10^3\,\,({\rm GeV/c})^2$ and 
 BW cut = $\Lambda_1$ = $0.2600\times 10^2$ GeV/c.
 QCD threshold = $\sqrt{t_{Q}} = 0.2060\times 10^2\,\, {\rm GeV/c}$.
 We take the number of flavor as $n_f= 3$ and the QCD scale parameter
  $\Lambda^{QCD} =0.213$ GeV. \\

 The calculated results are illustrated in Figs.1-4;  In Figs.1, 2 we give the 
results for the space-like momentum for $G_M^p/\mu_p$, $G_E^p$, $G_M^n/\mu_n$,
 $G_E^n$ and in Fig.3 the ratio of electric and magnetic form factors of proton  $\mu_pG_E^p/G_M^p$. In Fig.4 the results for
  the time-like momentum are illustrated for the proton and neutron form factors $|G^p|$ and $|G^n|$, respectively. Experimental data are taken from 
  \cite{bostedGMp}-\cite{antonelliGntm}. 
  
\begin{figure}
\begin{center}
\begin{tabular}{cc}
\includegraphics[width=.47\linewidth]{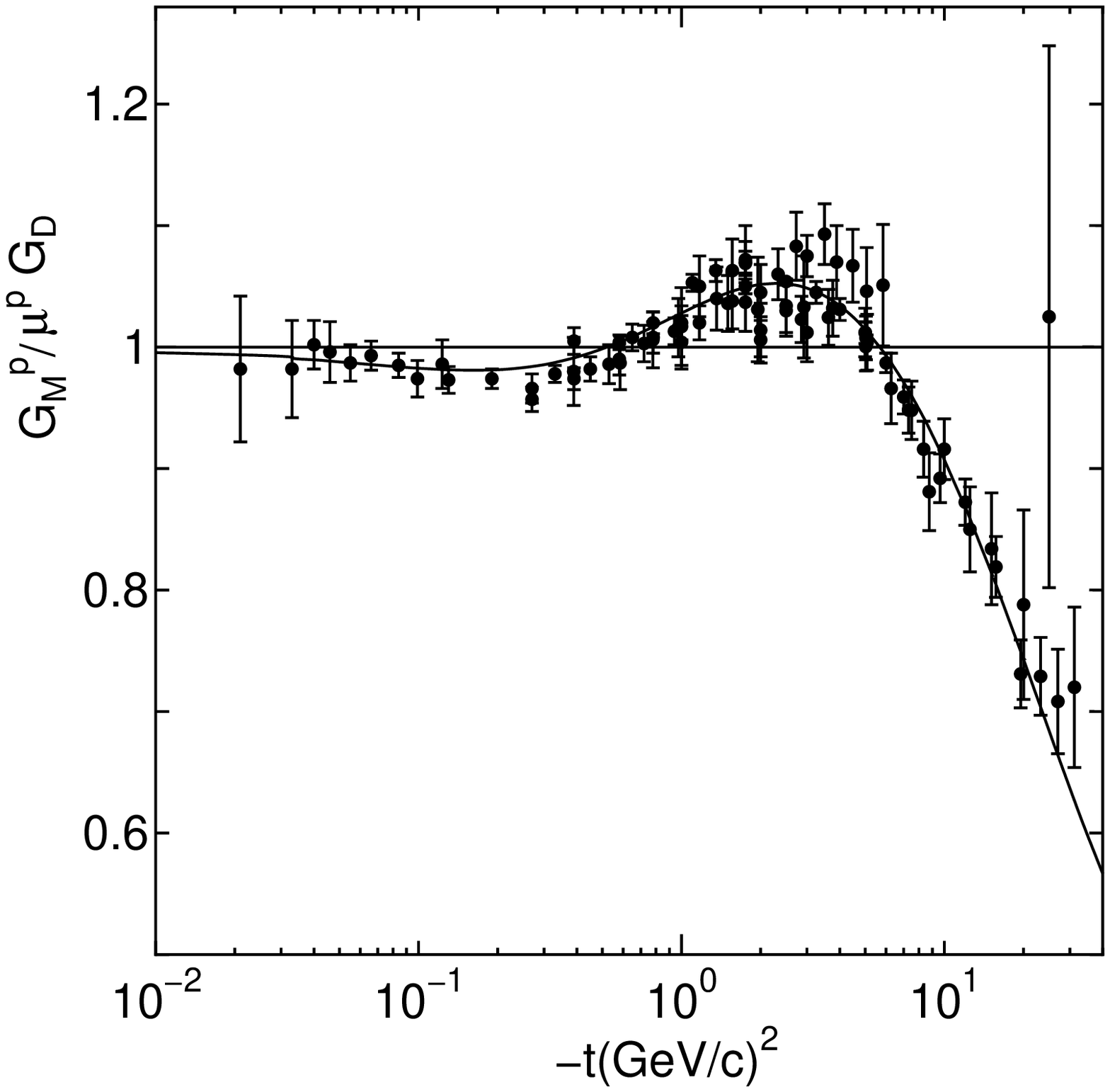}
& \includegraphics[width=.47\linewidth]{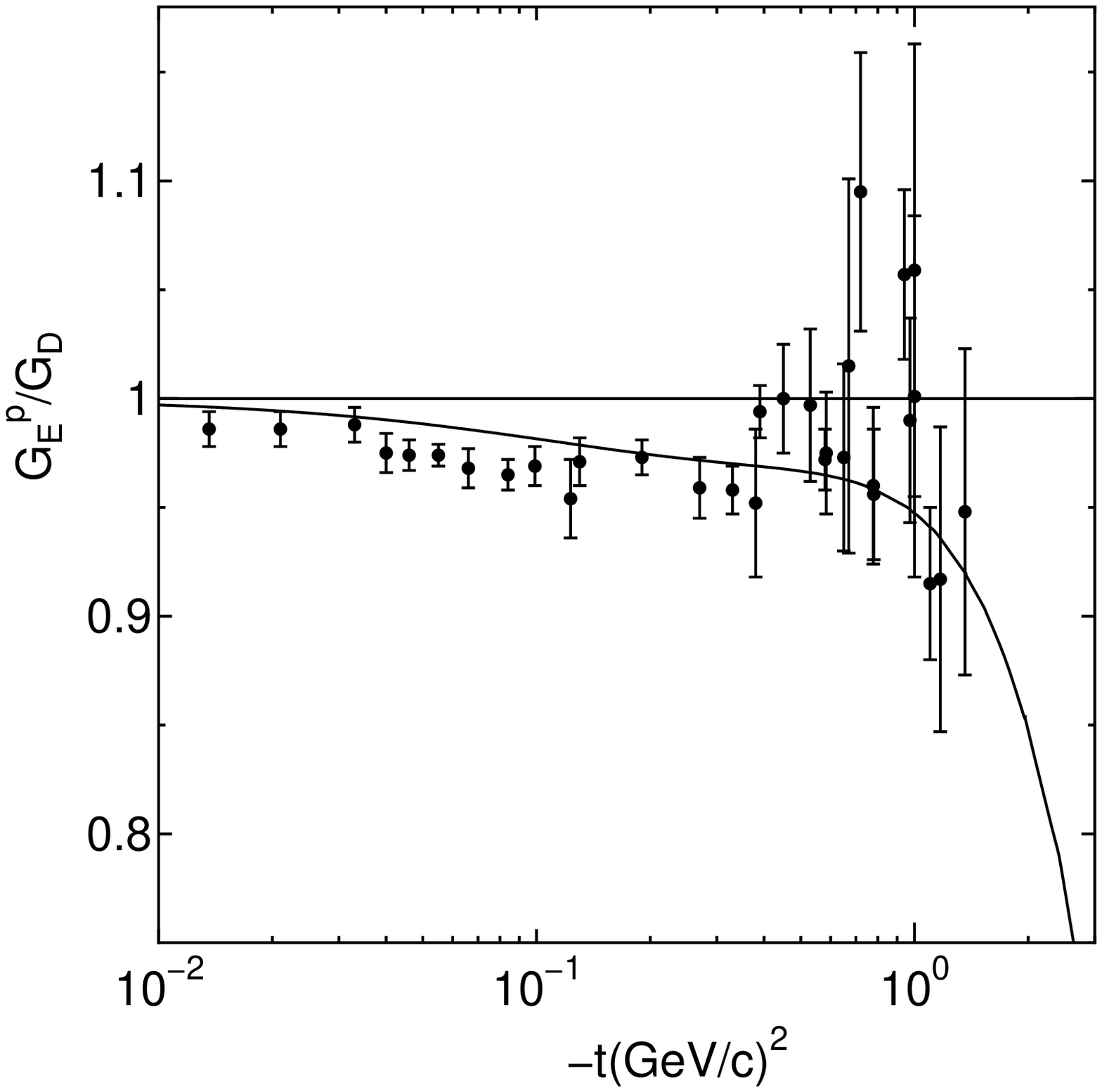} \\
\end{tabular}
\end{center}
Fig.1 Proton form factor for the space-like momentum: 
Magnetic and electric form factors.\\ 
\end{figure}
\begin{figure}
\begin{center}
\begin{tabular}{cc}
 \includegraphics[width=.47\linewidth]{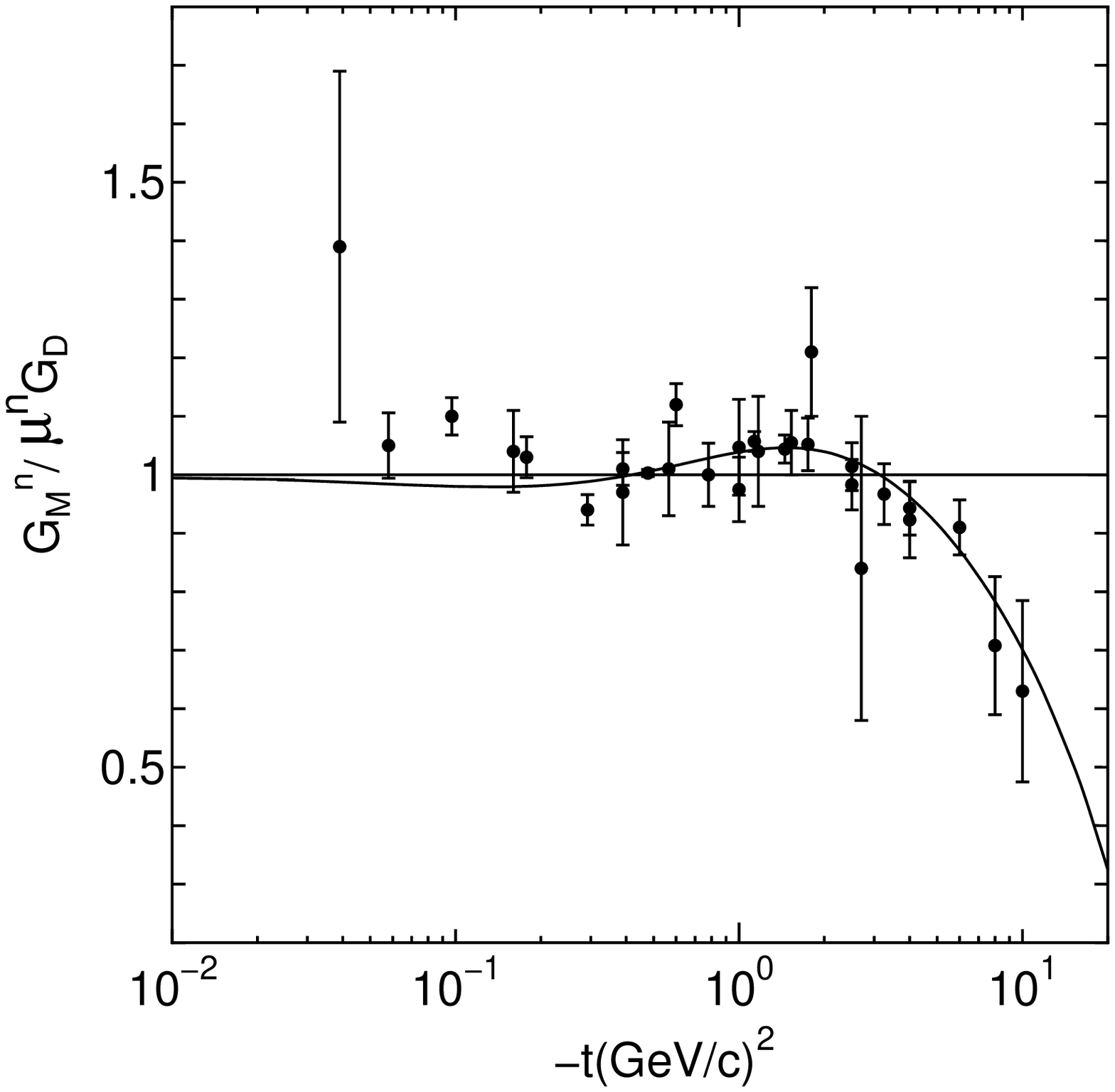}
& \includegraphics[width=.47\linewidth]{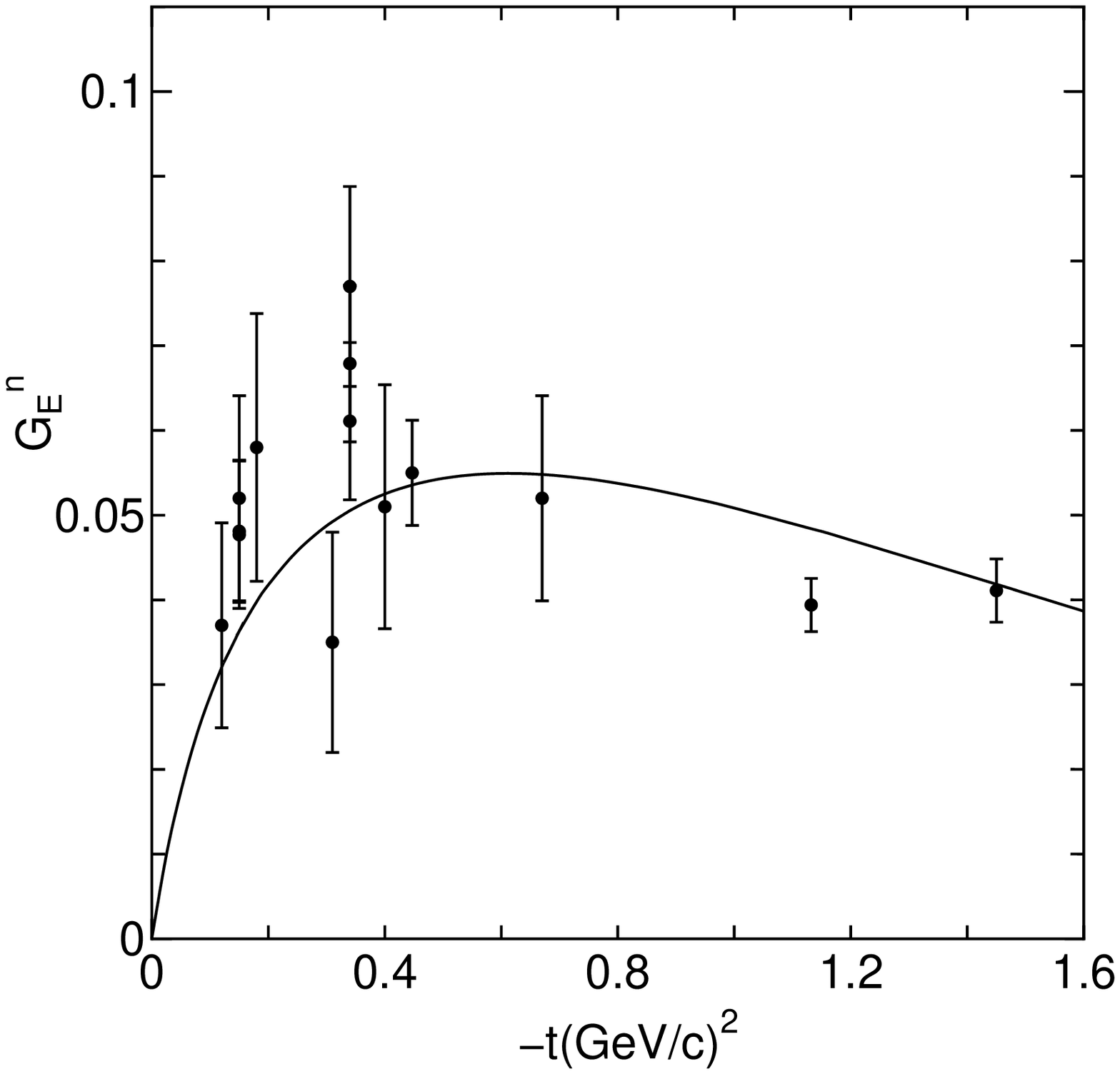} \\
\end{tabular}
\end{center}
Fig.2 Neutron form factors for the space-like momentum: 
Magnetic and electric form factors.
\end{figure}
\begin{figure}
\begin{center}
\begin{tabular}{cc}
\includegraphics[width=.47\linewidth]{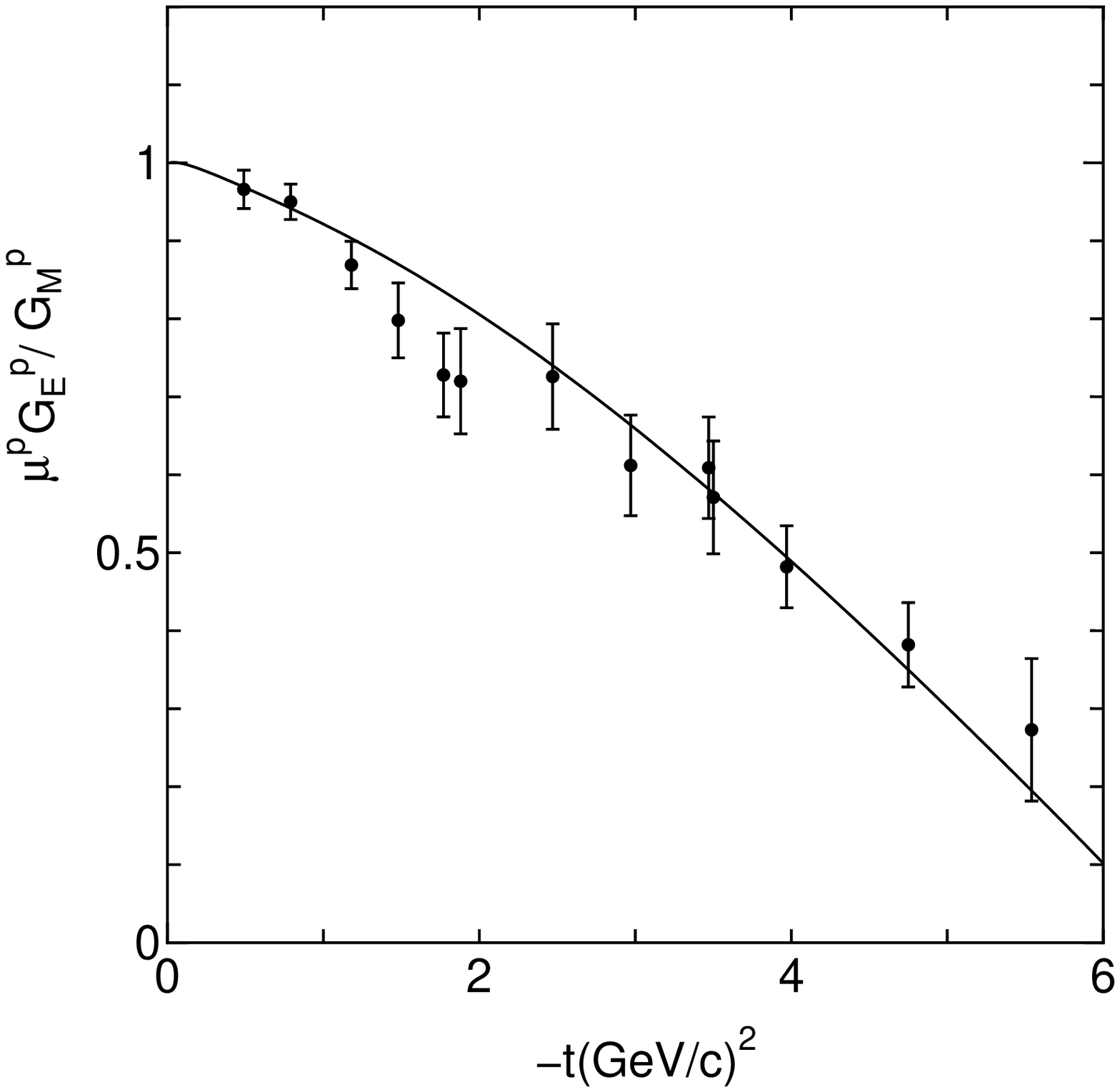}
& {} \\
\end{tabular}
\end{center}
Fig.3 Ratio of the electric and magnetic form factors of proton for the 
 space-like momentum.  
\end{figure}
\begin{figure}
\begin{center}
\begin{tabular}{cc}
\includegraphics[width=.47\linewidth]{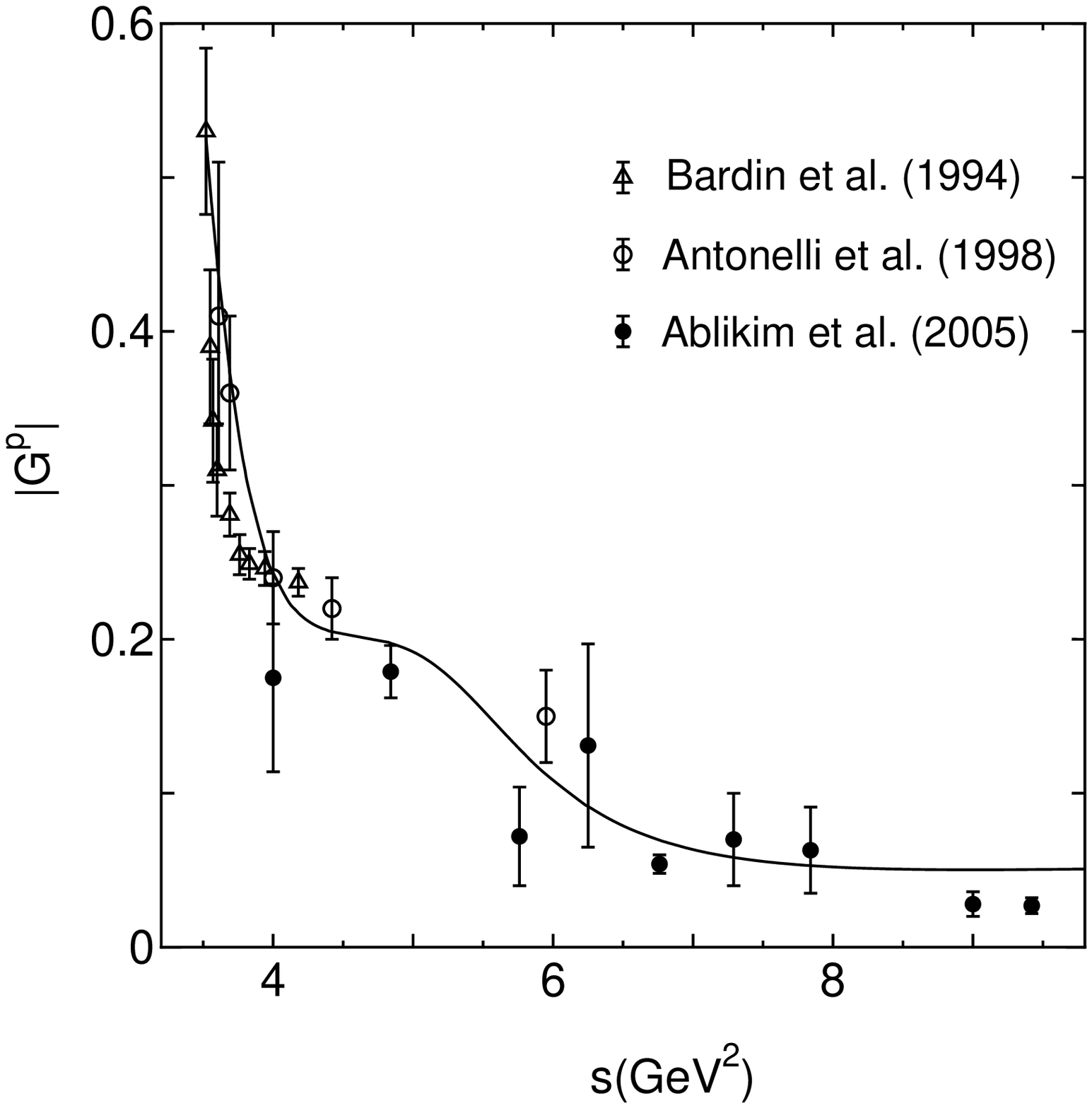}
& \includegraphics[width=.47\linewidth]{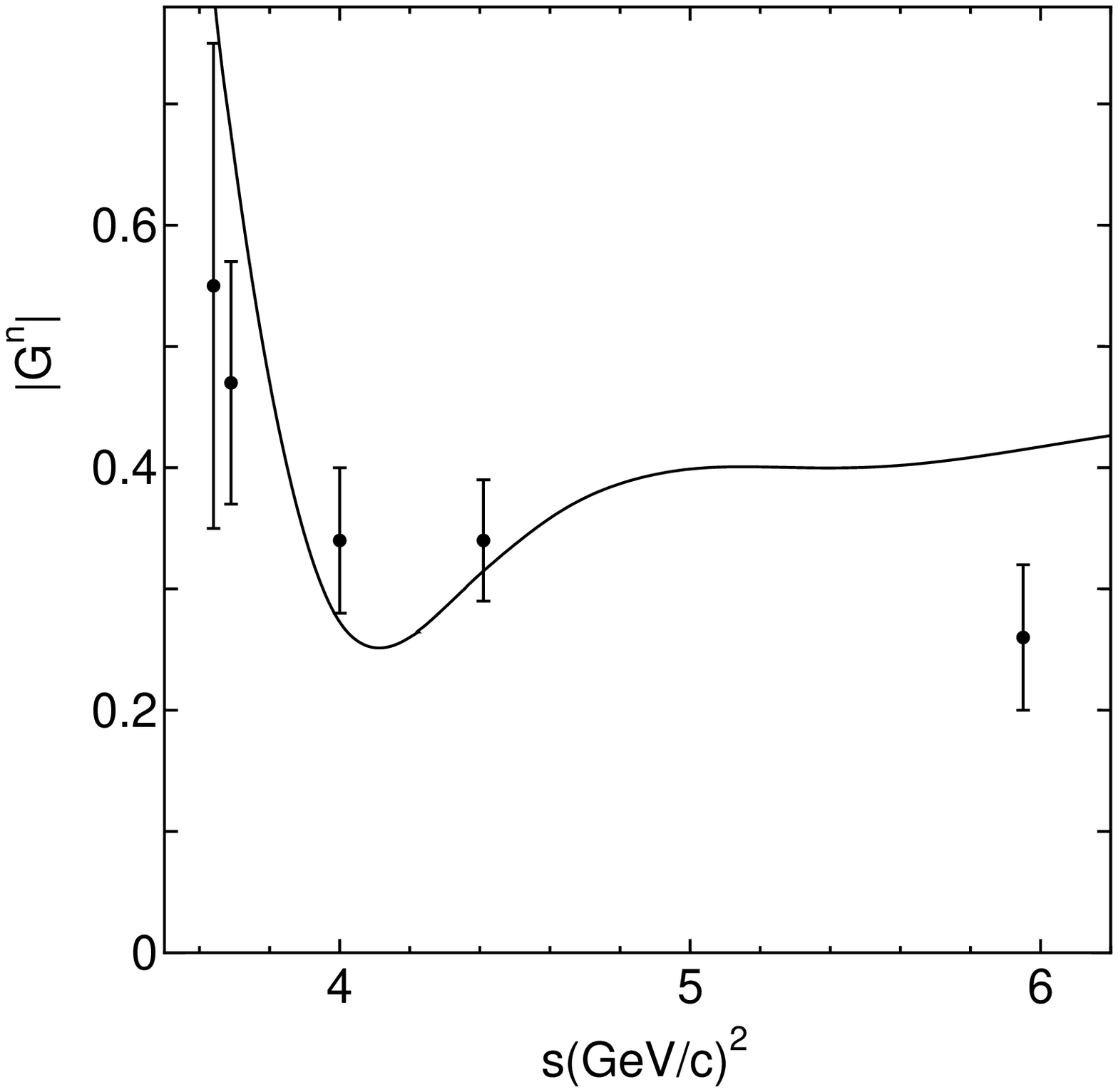} \\
\end{tabular}
\end{center}
Fig.4 Nucleon form factors for the time-like momentum: the proton and 
neutron form factors.
\end{figure}

\newpage
\section{Concluding remarks}

We have demonstrated that our superconvergent dispersion relation works 
in synthesizing the low and the high momentum parts of nucleon 
electromagnetic form factors for the space-like and time-like momentums  as we did for the bosons.

For the space-like momentum we were able to reproduce the experimental data, but for the time-like momentum we did not have very good results. 
If we restrict ourselves only to the data of space-like momentum, 
leaving out the time-like data in the chi square analysis, the result for the
 space-like momentum is improved a little; we have $\chi^2_{\rm space}
  = 217$.  By using the parameters 
 thus determined, we calculated the time-like part $|G|$, which
  turned out to be very large; the value of chi square became as large as
 $\chi^2_{\rm time} = 1.4 \times 10^6$. Incorporation of the data for the 
time-like momentum seems to be necessary in the systematic study of 
space-like and time-like momentum, although the number of data is limited. 

We used the experimental data for the helicity amplitudes obtained by 
H\"oher and Schopper in which the contribution from the $\rho$ meson 
is included. As their data are limited to low $t\,\, (\le 0.779$(GeV/c)$^2$),
 we do not have sufficient 
data for the region $s \le 4m_N^2$. We 
supplemented the unphysical region for $I=1$ state by introducing vector bosons
 with the small mass, $m_V \stackrel{\large <}{_{\sim}}1.4$ GeV/c$^2$.   For the isoscalar state 
we also introduced a vector boson with small mass. 

In our calculation we treated all of the vector boson masses and widths as 
parameters. If they are kept 
at experimental values, we get poor results. The superconvergence 
conditions are so strong  that the value of $\chi^2$ is very sensitive
 to the mass and width. The masses are obtained to be smaller 
than the experimental value and the existence of vector bosons with the masses 
around 1.2 $\sim$ 1.4 GeV/c$^2$ are implied.

To conclude the paper we remark on the  mass about 1.2 GeV/c$^2$.
  Both for $I=0$ and $I=1$ states there are indications of resonances observed 
  by the processes $e^{+}e^{-}\to \eta \pi^{+}\pi^{-}$, 
   $\gamma p \to \omega \pi^{0} p$ and $B \to D^{*}\omega\pi^{-}$ \cite{komada}. Incorporation of further resonances may improve results for 
   the time-like momentum.  \\

The authors wish to express gratitude to Professor M. Ishida for the valuable
 discussions  and comments. We also would like to thank Professor T. Komada for the 
 information on the 
vector bosons  with the mass around 1.2 GeV/c$^2$.

\end{document}